\documentclass[12pt]{iopart}

\usepackage{iopams}
\usepackage{setstack}
\usepackage{graphicx}
\usepackage{amsfonts,amssymb}

\usepackage{latexsym}
\usepackage{bm}

\newtheorem{defi}{Definition}[section]
\newtheorem{prop}[defi]{Proposition}
\newtheorem{theorem}[defi]{Theorem}
\newtheorem{conj}[defi]{Conjecture}
\newcommand{\beconj}{\begin{conj}}
\newcommand{\enconj}{\end{conj}}
\newcommand{\betheo}{\begin{theorem}}
\newcommand{\entheo}{\end{theorem}}
\newtheorem{condi}[defi]{Condition}
\newcommand{\becondi}{\begin{condi}}
\newcommand{\encondi}{\end{condi}}
\newcommand{\beprop}{\begin{prop}}
\newcommand{\enprop}{\end{prop}}

\newcommand{\bq}{\begin{eqnarray}}
\newcommand{\eq}{\end{eqnarray}}
\newcommand{\bitm}{\begin{itemize}}
\newcommand{\eitm}{\end{itemize}}

\def\E{\mathcal{E}}

\newcommand{\ket}[1]{\left\vert\kern.3ex#1\kern.3ex\right\rangle}
\newcommand{\bra}[1]{\left\langle\kern.3ex #1 \kern.3ex \right\vert}
\newcommand{\scalar}[2]{\langle\kern.3ex #1 \kern.3ex|\kern.3ex#2\kern.3ex\rangle}

\newcommand{\ie}{{\emph{i.e.}}}
\newcommand{\eg}{{\emph{e.g.}}}

\begin{document}
\noindent \hfill
MaPhy-AvH/2012-02\\

\title[New SUSYQM coherent states for P\"oschl-Teller potentials]{New SUSYQM coherent states for P\"oschl-Teller potentials: a detailed mathematical analysis}
\author{
H Bergeron$^{1}$ \footnote{herve.bergeron@u-psud.fr} , 
P Siegl$^{2,3,4}$ \footnote{siegl@ujf.cas.cz}
A Youssef $^{2,5}$ \footnote{youssef@mathematik.hu-berlin.de}} 

\address{$^1$ \textit{Universit\'e Paris-Sud, ISMO, UMR 8214 du CNRS, , B\^at. 351 F-91405 Orsay, France}\\
$^2$ \textit{Universit\'e Paris Diderot Paris 7, Laboratoire APC, Case 7020, F-75205 Paris}\\
$^3$ \textit{Czech Technical University in Prague, Faculty of Nuclear Sciences and Physical Engineering, B\v rehov\'a 7, 11519 Prague, Czech Republic } \\
$^4$ \textit{Nuclear Physics Institute ASCR, 25068 \v Re\v z, Czech Republic}\\
$^5$ \textit{Institut f\"ur Mathematik und Physik, Humboldt-Universit\"at zu Berlin, 12489 Berlin, Germany}\\}

\begin{abstract}
In a recent short note [Bergeron H, Gazeau J P, Siegl P and Youssef A 2010 {\it EPL} {\bf 92}  60003], we have presented the nice properties of a new family of semi-classical states for P\"oschl-Teller potentials. These states are built from a supersymmetric quantum mechanics approach and the parameters of these ``coherent" states are points in the classical phase space. In this article we develop all the mathematical aspects  that have been left apart in the previous article (proof of the resolution of unity, detailed calculations of quantized version of classical observables and mathematical study of the resulting operators: problems of domains, self-adjointness or self-adjoint extensions). Some additional questions as asymptotic behavior are also studied. Moreover, the framework is extended to a larger class of P\"oschl-Teller potentials.
  \end{abstract}
  \submitto{\JPA}
\date{\today} 

\maketitle

\section{Introduction}

Infinite wells are often modelled by P\"oschl-Teller (also known as trigonometric Rosen-Morse) confining potentials \cite{PT,RM} used, \eg~ in quantum optics \cite{Tomak,Wang}. The infinite square well is a limit case of this family. The question  is to find  a family of normalized states: (a) phase-space labelled, (b1) yielding a resolution of the identity, (b2) the latter holding with respect to  {\it the usual uniform measure on phase space}, (c) allowing a reasonable classical-quantum correspondence (``CS" quantization) and  (d) exhibiting semi-classical  phase space properties with respect to P\"oschl-Teller Hamiltonian time evolution. We refer to these states as coherent states (CS) as they share many striking properties with Schr\"odinger's original semi-classical states.

Most of the CS encountered in the literature are built through a group-theoretical or algebraic approach. Regarding  P\"oschl-Teller potentials, they  belong to the class of shape invariant potentials \cite{Gend} that have been intensively studied either specifically within the framework of supersymmetric quantum mechanics (SUSYQM) \cite{Book} or using a pure algebraic approach \cite{Bala1,Fukui}. Then various semi-classical states adapted to supersymmetric systems in general \cite{Fukui,Fatyga,Shreecharan} or to P\"oschl-Teller potentials in particular have been proposed in previous  works (see \cite{Crawford,Alex,Gaz,Kinani,Fernandez,Cruz} and references therein). \\
Whereas most of them verify (b1) and (d), they do not really ``live" on the genuine classical phase space of the system. Hence, a classical-quantum correspondence (property (c))  often lacks unambiguous interpretation. Moreover, the correspondence  between  classical and quantum  momenta for a particle moving on an interval requires a thorough analysis; as a matter of fact, there exists  a well-known ambiguity  in the definition of the quantum momentum operator \cite{reedsimon,Gaz}. 

In a recent note \cite{HJP2010}, we have presented a construction of coherent states for  P\"oschl-Teller potentials based on a general approach given  by one of us in \cite{Herve}, and we have displayed their remarkable qualities as classical-quantum ``conveyers''. These states fulfil the conditions (a), (b1), (b2), (c) and (d). The property (b2) is specially unexpected because  non-linear CS verify in general a resolution of unity with respect to some positive weight function. The validity of (b2) means that our CS do not favour any part of the classical phase space, even if this phase space is a strip (P\"oschl-Teller potentials case), that is a manifold with boundaries, topologically very different from the whole plane of the usual (harmonic) CS.

In this article we examine in details the mathematical aspects of properties (a), (b1), (b2) and (c) as well as some additional questions. In particular we pay more attention at the ``quantization procedure" (c), analyzing in details all the mathematical subtleties due to the unbounded character of most operators (domains, closure, possibly unique (or not) self-adjoint extensions). Furthermore, due to their applications in quantum dots and quantum wells, only symmetric repulsive P\"oschl-Teller potentials have been considered in our note \cite{HJP2010}. But in fact our formalism remains valid for a larger class of P\"oschl-Teller potentials that is considered in the following.

\section{The P\"oschl-Teller Hamiltonian and SUSYQM formalism}

\subsection{The P\"oschl-Teller Hamiltonian}

We consider the quantum problem of a particle trapped on the interval $[0,L]$. The Hilbert space is $\mathcal{H}=L^2([0,L],dx)$ and the Hamiltonian is the following Sturm-Liouville operator ${\bf H}_{\nu,\beta}$ (self-adjoint when defined on a suitable dense domain $\mathcal{D}_{{\bf H}_{\nu,\beta}}$ of $\mathcal{H}$ that will be specified in the next section)
\begin{equation}
\label{equa:poschlt}
{\bf H}_{\nu,\beta}=-\frac{\hbar^2}{2m}\frac{d^2}{dx^2}+V_{\E_0,\nu,\beta}(x).
\end{equation}
$V_{\nu,\beta}$ is the  P\"oschl-Teller potential 
 \begin{equation}
 V_{\E_0,\nu,\beta}(x)=\frac{\E_0 \nu(\nu+1)}{\sin^2 \frac{\pi x}{L}}-2 \E_0 \beta \cot \frac{\pi x}{L}
 \end{equation}
where  $\E_0$ is some energy scale, while $\nu$ and $\beta$ are some dimensionless parameters. We restrict our study to the repulsive behavior at the end points $x=0$ and $x=L$. This assumption allows us to choose $\E_0 \ge 0$ and $\nu \ge 0$. Moreover since the symmetry $x \mapsto L-x$ corresponds to the parameter change $\beta \mapsto -\beta$, we can freely choose $\beta \ge 0$.
 
Now, since the potential strengths are overdetermined by specifying $\E_0$, $\nu$ and  $\beta$, we can freely choose the energy scale $\E_0$ as the zero point energy of the infinite well, namely
 $ \E_0=\hbar^2 \pi^2/(2m L^2)$. Then $\nu$ and $\beta$ remain the unique free parameters of the problem. {\it In the sequel $\nu$ and $\beta$ are always assumed to be  positive except if a contrary assumption is specified}.\\
The case $\beta=0$ corresponds to the symmetric repulsive potentials investigated in \cite{HJP2010}, while the case $\beta \ne 0$ leads to the Coulomb potential in the limit $L \to \infty$ (if we choose $\beta=Z e^2 m L/(4 \pi^2 \epsilon_0 \hbar^2)$).

\subsection{Functional point of view and self-adjointness}

The P\"oschl-Teller Hamiltonian \eref{equa:poschlt} is an ordinary differential Sturm-Liouville operator, singular at end points. The functional properties depend on the value of $\nu$, as follows from the analysis of Gesztesy {\it et al.} \cite{Gesztesy}. In particular $\nu=1/2$ is the critical value, while one would naively expect $\nu=0$, \ie~the infinite square well, to play the role.
 
Let us define operator ${\bf \dot{H}}_{\nu,\beta}$ with the action given by the formal differential expression $\tau \psi = -\hbar^2/(2m) \, \psi''+V_{\E_0,\nu,\beta}(x)\psi$
and with the domain $\mathcal{C}_0^\infty(0,L)$, \ie~smooth functions with a compact support. Using the standard approach and terminology \cite{reedsimon2}, the P\"oschl-Teller potential $V_{\nu,\beta}$ is in the limit point case at both ends $x=0$ and $x=L$, if $\nu \ge 1/2$, and in the limit circle case at both ends if $0 \le \nu < 1/2$. It follows that ${\bf \dot{H}}_{\nu,\beta}$ is essentially self-adjoint in the former case. The closure is denoted ${\bf H}_{\nu,\beta}$ and its domain coincides with the maximal one, \ie~
$\mathcal{D}_{{\bf H}_{\nu,\beta}}=\{\psi \in ac^2(0,L) \, | \, \tau \psi \in \mathcal{H}\}$,
where $ac^2(0,L)$ denotes absolutely continuous functions with absolutely continuous derivatives.
It is possible to check that the function from this domain automatically satisfy Dirichlet boundary conditions.
In the latter range of $\nu$, deficiency indices of ${\bf \dot{H}}_{\nu,\beta}$ are $(2,2)$ and therefore more self-adjoint extensions exist; see \cite{Gesztesy} for the detailed analysis. In this paper we select the extension described by Dirichlet boundaries conditions, \ie~
$\mathcal{D}_{{\bf H}_{\nu,\beta}}=\{\psi \in ac^2(0,L) \, | \, \psi(0)=\psi(L)=0,\, \tau \psi \in \mathcal{H}\}$.
For further use, we define the dense domain, being the common core for ${\bf H}_{\nu,\beta}$,
\begin{equation}
\label{equa:domH}
\mathcal{D}_H=\{\psi \in AC^2(0,L) \, | \, V_{\E_0,\nu,\beta} \psi  \in \mathcal{H} \}.
\end{equation}
where $AC(0,L) = \{\psi \in ac(0,L) \, | \, \psi' \in \mathcal{H} \}$ and $AC^2(0,L)$ is introduced analogously.

 \subsection{Eigenvalues and eigenfunctions}

The eigenvalue problem is explicitly solvable; the eigenvalues $E_n^{(\nu,\beta)}$ and corresponding eigenfunctions $\phi_n^{(\nu,\beta)}$ read
\begin{eqnarray}
E_n^{(\nu,\beta)} &= \E_0 \left( (n+\nu+1)^2 - \frac{\beta^2}{(n+\nu+1)^2} \right) \\
\phi_n^{(\nu,\beta)}(x) &=  K_n^{(\nu,\beta)} \sin^{\nu+n+1} \Big( \frac{\pi x}{L} \Big) 
\exp \Big( \frac{\beta \pi x}{L(\nu+n+1)} \Big)
P_n^{(a_n,\bar{a}_n)} \Big( i \cot \frac{\pi x}{L} \Big)
\end{eqnarray}
where $n \in \mathbb{N}_0$, $a_n=-(\nu+n+1-i \beta (\nu+n+1)^{-1})$, $K_n^{(\nu,\beta)}$ is a normalization constant, and the $P_n^{(a,b)}$ are the Jacobi polynomials. An expression using only real polynomials can be found in ~\cite{Compean2008}. The ground state $\phi_0^{(\nu,\beta)}$ simplifies to
 \begin{equation}
 \phi_0^{(\nu,\beta)}(x)= K_0^{(\nu,\beta)} \sin^{\nu+1} \left( \frac{\pi x}{L} \right) \exp \left( \frac{\beta \pi x}{L(\nu+1)}\right)
 \end{equation}
and the eigenfunctions $\phi_n^{(\nu,0)}$ for $\beta=0$ can be also expressed in terms of Gegenbauer polynomials $\mathrm{C}_n^{\nu+1}$ as:
 \begin{equation}
 \phi_n^{(\nu,0)}(x)=Z_{n,\nu} \sin^{\nu+1} \left( \frac{\pi x}{L} \right)  
\mathrm{C}_n^{\nu+1} \left( \cos \left(\frac{\pi x}{L} \right)\right)
 \end{equation}
where $Z_{n,\nu}$ is a normalization constant. Finally, the eigenfunctions for the infinite well ($\nu=\beta=0$) read
 \begin{equation}
 \phi_n^{(0,0)}(x)=\sqrt{\frac{2}{L}} \sin \left( \frac{(n+1)\pi x}{L}\right).
 \end{equation}

\subsection{SUSYQM and Shape Invariance of P\"oschl-Teller Hamiltonians}

We use a standard SUSY approach, leading to a simple Darboux factorization of the Hamiltonian (for more details about SUSY and factorization problems see \cite{Mielnik,Andrianov}). The superpotential $W_{\nu,\beta}(x)$ can be found as
\begin{equation}
W_{\nu,\beta}(x)=-\hbar \frac{(\phi_0^{(\nu,\beta)})'(x)}{\phi_0^{(\nu,\beta)}(x)}=-\frac{\hbar \pi}{L} \left( (\nu+1) \cot \frac{\pi x}{L} - \frac{\beta}{\nu+1} \right).
\end{equation}
We define the operators ${\bf A}_{\nu,\beta}$ and ${\bf A}_{\nu,\beta}^{\dagger}$ as the differential operators
\begin{equation}
{\bf A}_{\nu,\beta} = W_{\nu,\beta}(x)+ \hbar \frac{d}{dx} \quad  {\rm and } \quad {\bf A}^{\dagger}_{\nu,\beta} = W_{\nu,\beta}(x)-\hbar \frac{d}{dx}
\end{equation}
acting in the domains 
\begin{eqnarray}
\label{equa:domA}
\mathcal{D}_{{\bf A}_{\nu,\beta}} &= \{ \psi \in ac(0,L) \, | {\bf A}_{\nu,\beta} \psi \in \mathcal{H} \}  
\quad  {\rm and } \quad \nonumber \\
\mathcal{D}_{{\bf A}_{\nu,\beta}^{\dagger}}& = \{ \psi \in ac(0,L) \, | {\bf A}_{\nu,\beta}^{\dagger} \psi \in \mathcal{H} \}. 
\end{eqnarray}

It can be verified that ${\bf A}_{\nu,\beta}^{\dagger}$ is indeed the adjoint of ${\bf A}_{\nu,\beta}$ and that functions from domains (\ref{equa:domA}) satisfy Dirichlet boundary conditions. Besides the domains $\mathcal{D}_{{\bf A}_{\nu,\beta}}$ and $\mathcal{D}_{{\bf A}_{\nu,\beta}^{\dagger}}$, we consider their common restriction
\begin{equation}\label{equa:domD}
\mathcal{D}_{A} = \{ \psi \in AC(0,L) \, | \, W_{\nu,\beta} \psi \in \mathcal{H} \}.
\end{equation}
It can be verified that 
$\overline{{\bf A}_{\nu,\beta}{\upharpoonright \mathcal{D}_A}} = {\bf A}_{\nu,\beta}$ and 
$\overline{{\bf A}_{\nu,\beta}^{\dagger}{\upharpoonright \mathcal{D}_A}} = {\bf A}_{\nu,\beta}^{\dagger}$ and moreover, it seems that domain $\mathcal{D}_A$ is a suitable choice in the CS quantization procedure, see Section \ref{sec:quant}.\\
${\bf A}_{\nu,\beta}$ and ${\bf A}_{\nu,\beta}^{\dagger}$ are not ladder operators. As shown in (\ref{equa:eigenstates}), these operators connect the eigenvectors of the supersymmetric partner Hamiltonians ${\bf H}^{(S)}_{\nu,\beta}$ and ${\bf H}_{\nu,\beta}$ defined below. Only in the case of the harmonic potential, the corresponding differential operators ${\bf A}$ and ${\bf A}^\dag$ lead to usual lowering and raising operators.\\
Nevertheless, while being not a ladder operator,  ${\bf A}_{\nu,\beta}$ allows to build a family of ``coherent states'' (following our CS definition of the beginning of the introduction) that possess very interesting properties \cite{HJP2010, Herve}.

The P\"oschl-Teller Hamiltonian ${\bf H}_{\nu,\beta}$ can be factorized with help of ${\bf A}_{\nu,\beta}$ and ${\bf A}_{\nu,\beta}^{\dagger}$ as:
\begin{equation}
{\bf H}_{\nu,\beta}=\frac{1}{2m}{\bf A}^\dag_{\nu,\beta} {\bf A}_{\nu,\beta} + E_0^{(\nu,\beta)}. 
\end{equation}
where this equality holds in operator sense as well.
The supersymmetric partner ${\bf H}^{(S)}_{\nu,\beta}$ of ${\bf H}_{\nu,\beta}$ is defined as:
\begin{equation}
{\bf H}^{(S)}_{\nu,\beta}=\frac{1}{2m} {\bf A}_{\nu,\beta} {\bf A}^\dag_{\nu,\beta} + E_0^{(\nu,\beta)}. 
\end{equation}
and by simple manipulations we find
\begin{equation}
{\bf H}^{(S)}_{\nu,\beta}= {\bf H}_{\nu+1,\beta}.
\end{equation}
This relation specifies that P\"oschl-Teller Hamiltonians are shape invariant.
From the general features of supersymmetric partner Hamiltonians, if we call $\tilde{E}_n^{(\nu,\beta)}$ the eigenvalues of ${\bf H}^{(S)}_{\nu,\beta}$ and $\ket{\tilde{\phi}_n^{(\nu,\beta)}}$ the corresponding eigenstates, we have
\begin{equation}
\label{equa:eigenstates}
\tilde{E}_n^{(\nu,\beta)}=E_{n+1}^{(\nu,\beta)} \quad {\rm and } \quad \ket{\tilde{\phi}_n^{(\nu,\beta)}}=\frac{1}{\sqrt{2m(E_{n+1}^{(\nu,\beta)}-E_0^{(\nu,\beta)})}} {\bf A}_{\nu,\beta} \ket{\phi_{n+1}^{(\nu,\beta)}}.
\end{equation}
If we introduce the positive sequence $f_n^{(\nu,\beta)}=\left( E_n^{(\nu,\beta)}-E_0^{(\nu,\beta)} \right) \E_0^{-1}$, then  ${\bf A}_{\nu,\beta}$ and ${\bf A}_{\nu,\beta}^{\dagger}$ can be decomposed as
\begin{equation}
\label{AA:dec}
\left\{
\begin{array}{c}
{\bf A}_{\nu,\beta}=\sqrt{2m \E_0} \sum_{n=0}^\infty \sqrt{f_{n+1}^{(\nu,\beta)}} \ket{\tilde{\phi}_n^{(\nu,\beta)}} \bra{\phi_{n+1}^{(\nu,\beta)}} \\ 
{\bf A}^\dag_{\nu,\beta} = \sqrt{2m \E_0} \sum_{n=0}^\infty \sqrt{f_{n+1}^{(\nu,\beta)}} \ket{\phi_{n+1}^{(\nu,\beta)}} \bra{\tilde{\phi}_n^{(\nu,\beta)}}
\end{array}.
\right.
\end{equation}
Furthermore, since ${\bf H}^{(S)}_{\nu,\beta}={\bf H}_{\nu+1,\beta}$, we have $\tilde{E}_n^{(\nu,\beta)}=E_n^{(\nu+1,\beta)}$ and $\ket{\tilde{\phi}_n^{(\nu,\beta)}}=\ket{\phi_n^{(\nu+1,\beta)}}$ and we deduce easily the recurrence relation 
\begin{equation}
E_n^{(\nu,\beta)}=E_{n-1}^{(\nu+1,\beta)}= \dots= E_0^{(\nu+n,\beta)}.
\end{equation}
Using the latter we can simplify the expressions (\ref{AA:dec}) to
\begin{equation}
\left\{
\begin{array}{c}
{\bf A}_{\nu,\beta}=\sqrt{2m \E_0} \sum_{n=0}^\infty \sqrt{f_{n+1}^{(\nu,\beta)}} \ket{\phi_n^{(\nu+1,\beta)}} \bra{\phi_{n+1}^{(\nu,\beta)}} \\ 
{\bf A}^\dag_{\nu,\beta} = \sqrt{2m \E_0} \sum_{n=0}^\infty \sqrt{f_{n+1}^{(\nu,\beta)}} \ket{\phi_{n+1}^{(\nu,\beta)}} \bra{\phi_n^{(\nu+1,\beta)}}
\end{array}.
\right.
\end{equation}
and obtain a rule how to construct the eigenstates $\ket{\phi_{n+1}^{(\nu,\beta)}}$ from the ground state:
\begin{equation}
\fl
 \ket{\phi_{n+1}^{(\nu,\beta)}}=\left( \frac{1}{\sqrt{2m\E_0}} \right) ^{n+1} \frac{1}{\sqrt{f_{n+1}^{(\nu,\beta)} f_n^{(\nu+1,\beta)} \dots f_1^{(\nu+n,\beta)}} } {\bf A}_{\nu,\beta}^{\dag} {\bf A}_{\nu+1,\beta}^{\dag} \dots {\bf A}_{\nu+n,\beta}^{\dag} \ket{\phi_0^{(\nu+n+1,\beta)}}.
\end{equation}

\section{The coherent states and their properties}

\subsection{The coherent states}

We define the coherent state $\ket{\xi_z^{[\nu,\beta]}}$,  $z \in \mathbb{C}$, as the eigenstate of ${\bf A}_{\nu,\beta}$ associated to the eigenvalue $z$. Up to a normalization factor, we obtain
\begin{equation}
\xi_z^{[\nu,\beta]}(x)= \sin^{\nu+1} \frac{\pi x}{L} \exp \left( \frac{ z x}{\hbar} - \frac{\beta \pi x}{L(\nu+1)} \right) \quad {\rm for } \quad x \in [0,L].
\end{equation}
The set 
$\mathcal{K}=\{ (q,p)\, | \, q\in [0,L],  \, p\in \mathbb{R} \}$
corresponds to the classical phase space of the P\"oschl-Teller problem.
Inspired by the structure of operator ${\bf A}_{\nu,\beta}$ that reads (when restricted to $\mathcal{D}_A$)
${\bf A}_{\nu,\beta} =W_{\nu,\beta}({\bf Q})+i {\bf P}_\#$, where we introduced the operators ${\bf Q}: \psi(x) \to x\psi(x)$ and ${\bf P}_\#: \psi \to -i \hbar \psi'(x)$ (defined on $\mathcal{D}_A$), we change the variable $z=W_{\nu,\beta}(q)+ip$ into $\ket{\xi_{W_{\nu,\beta}(q)+ip}^{[\nu,\beta]}}$ with $0<q<L$ and $p \in \mathbb{R}$. Developing the exponential part of the state, we find that the $\beta$ dependence disappears 
%
$\ket{\xi_{W_{\nu,\beta}(q)+ip}^{[\nu,\beta]}}=\ket{\xi_{W_{\nu,0}(q)+ip}^{[\nu,0]}}$.
%
Hence we have received the following family of normalized coherent states $\eta_{q,p}^{[\nu]}$, independent of $\beta$, 
\begin{equation}
\ket{\eta_{q,p}^{[\nu]}}= N_{\nu}(q) \ket{\xi_{W_{\nu,0}(q)+ip}^{[\nu,0]}},
\end{equation}
where the normalization constant $N_{\nu}(q)$ reads
\begin{equation}
\frac{1}{N_{\nu}^2(q)}=\int_0^{L} |\xi_{W_{\nu,0}(q)+ip}^{[\nu,0]}(x)|^2 dx=\int_0^{L} \sin^{2\nu+2} \left( \frac{\pi x}{L} \right) e^{2 W_{\nu,0}(q) x/\hbar} dx.
\end{equation}
The Appendix contains the proof of the relation
\begin{equation}
\label{equa:intfornorm}
\fl 
\forall z  \in \mathbb{C}, \  \forall \nu>-3/2, \int_0^1 \sin^{2\nu+2}( \pi x ) e^{z x} dx= \frac{\Gamma(2\nu+3) e^{z/2}}{4^{\nu+1} \Gamma(\nu+2+i \frac{z}{2\pi}) \Gamma(\nu+2-i \frac{z}{2\pi})},
\end{equation}
and in the sequel we use the notation
\begin{equation}
\label{equa:defF}
 \forall z \in \mathbb{C} {\rm, } \ \forall \nu>-3/2 {\rm, } \quad F_{\nu}(z)=\int_0^1 \sin^{2\nu+2}( \pi x ) e^{z x} dx.
\end{equation}
After a simple change of variable, relation \eref{equa:intfornorm} yields that the normalization constant $N_{\nu}(q)$ can be expressed as
\begin{equation}
\label{equa:normalization}
\fl \ \ \ \ \ \ \ \
N_{\nu}(q)= \frac{1}{\sqrt{F_\nu(2W_{\nu,0}(q)L/\hbar)L}}=\frac{2^{\nu+1} |\Gamma(\nu+2+i (\nu+1) \lambda_q)|}{\sqrt{L} \sqrt{ \Gamma(2\nu+3)}} e^{-\frac{\pi}{2}(\nu+1) \lambda_q},
\end{equation}
where $\lambda_q= - \cot \frac{\pi q}{L}$.
%
The scalar product of two coherent states verifies
\begin{equation}
\scalar{\eta_{q,p}^{[\nu]}}{\eta_{q',p'}^{[\nu']}}=L \, N_{\nu}(q) N_{\nu'}(q') F_{\frac{\nu+\nu'}{2}} \left( \frac{L}{\hbar} \alpha \right)
\end{equation}
with $\alpha= W_{\nu,0}(q)+W_{\nu',0}(q')+i(p'-p)$.

\subsection{The resolution of unity}

The coherent states yield the following resolution of unity in week sense
\begin{equation}
\label{equa:resol}
\forall \nu \ge 0 {\rm, } \int_{\mathcal{K}} \frac{dqdp}{2 \pi \hbar} \ket{\eta_{q,p}^{[\nu]}} \bra{\eta_{q,p}^{[\nu]}} = \mathbb{I}.
\end{equation}
The proof is essentially based on the properties of the Fourier transformation. In addition we need the following integral calculated in the Appendix
\begin{equation}
\label{equa:intforresol}
\fl 
\forall x \in ]0,1[ {\rm, } \, \forall \nu > -1 {\rm, } \, \frac{4^{\nu}}{\pi^2} \int_{\mathbb{R}} \frac{ | \Gamma(\nu+1+i \frac{u}{2 \pi})|^2 e^{-u/2}}{\Gamma(2\nu+2)} e^{u x} du = \frac{1}{\sin^{2 \nu+2}(\pi x)}.
\end{equation}
\subsubsection{The proof}
Let $\psi \in \mathcal{H}$, the scalar product $\scalar{\eta_{q,p}^{[\nu]}}{\psi}$ reads, by definition, as
\begin{equation}
\scalar{\eta_{q,p}^{[\nu]}}{\psi}=N_{\nu}(q) \int_0^{L} \sin^{\nu+1} \frac{\pi x}{L} e^{\frac{W_{\nu,0}(q)}{\hbar}x} \psi(x) e^{-i \frac{p}{\hbar} x} dx.
\end{equation}
Let us define the function $f \in L^2(\mathbb{R},dx)$ as
\begin{equation}
f_q(x)=\mathbb{I}_{[0,L]}(x) \sin^{\nu+1} \frac{\pi x}{L} e^{\frac{W_{\nu,0}(q)}{\hbar}x} \psi(x),
\end{equation}
then $\scalar{\eta_{q,p}^{[\nu]}}{\psi}=N_{\nu}(q) \hat{f}_q(p/\hbar)$, where $\hat{f}_q$ stands for the Fourier transform of $f_q$. Since $f_q$ is at the same time an $L^1$ and $L^2$ function, the Plancherel-Parseval theorem yields:
\begin{equation}
\int_{\mathbb{R}} \frac{dp}{2\pi\hbar} |\scalar{\eta_{q,p}^{[\nu]}}{\psi}|^2=N_{\nu}(q)^2 \int_0^{L} |f_q(x)|^2 dx.
\end{equation}
Moreover, since we are manipulating with positive functions, the Fubini theorem can be used as well and
\begin{equation}
\int_{\mathcal{K}} \frac{dqdp}{2 \pi \hbar}  |\scalar{\eta_{q,p}^{[\nu]}}{\psi}|^2= \int_0^{L} dx  \int_0^{L} dq N_{\nu}(q)^2 |f_q(x)|^2.
\end{equation}
Finally, using the expression for $f_q(x)$ and the already mentioned integral relation, we obtain
\begin{equation}
\int_{\mathcal{K}} \frac{dqdp}{2 \pi \hbar}  |\scalar{\eta_{q,p}^{[\nu]}}{\psi}|^2= \int_0^{L} dx |\psi(x)|^2.
\end{equation}
The resolution of identity follows from the polarization identity.

\subsubsection{Remark}

The coherent states $\eta_{q,p}^{[\nu]}$ have been defined for $\nu \ge 0$. But in fact $\eta_{q,p}^{[\nu]} \in \mathcal{H}$ even for $\nu \ge -3/2$. Furthermore, the relation  \eref{equa:intforresol} holds for $\nu > -1$, consequently the resolution of unity can be extended from $\nu \ge 0$ to $\nu > -1$. This remark can be useful to extend some special formulae of quantized quantities studied in the next sections.

\subsection{Quantum frames and reproducing kernels in phase space}

The resolution of unity shown above proves that the kernels $K_\nu(p,q;p',q')=\scalar{\eta_{q,p}^{[\nu]}}{\eta_{q',p'}^{[\nu]}}$ are, in fact, reproducing kernels in $L^2(\mathcal{K},(2 \pi \hbar)^{-1}dqdp)$, which are similar to the well-known Fock-Bargmann-Segal reproducing kernel obtained with the usual harmonic coherent states. Then the kernel $K_\nu$ defines an orthogonal projector $\Pi_\nu$ acting on $L^2(\mathcal{K},(2 \pi \hbar)^{-1}dqdp)$. Let us call $\mathcal{H}_\nu={\rm Ran}(\Pi_\nu)$ the Hilbert subspace of $L^2(\mathcal{K},(2 \pi \hbar)^{-1}dqdp)$ associated with $\Pi_\nu$. Each family of functions $\{  \psi_n^{(\nu,\beta)} \}_{n \in \mathbb{N}}$ defined as $\psi_n^{(\nu,\beta)}(q,p)=\scalar{\eta_{q,p}^{[\nu]}}{\phi_n^{(\nu,\beta)}}$ is an orthonormal basis of $\mathcal{H}_\nu$ and defines a {\it quantum frame} in $L^2(\mathcal{K},(2 \pi \hbar)^{-1}dqdp)$. According to the general scheme \cite{alienglis2005,gazeau2009},  a Klauder-Berezin-Toeplitz quantization procedure is developed in the following section.

\section{Klauder-Berezin-Toeplitz quantization and some operator expressions}
\label{sec:quant}

\subsection{Preliminaries}

Taking into account the resolution of unity (\ref{equa:resol}) we quantize classical observables $f(q,p)$ defined on the phase space $\mathcal{K}$ by the correspondence
\begin{equation}
\label{equa:qmap}
f(q,p) \to {\bf F}=\int_{\mathcal{K}} \frac{dqdp}{2 \pi \hbar} f(q,p) \ket{\eta_{q,p}^{[\nu]}}\bra{\eta_{q,p}^{[\nu]}}
\end{equation}
where this integral is understood in the weak sense. This means that the integral defines in fact a sesquilinear form (eventually only densely defined)
\begin{equation}
\label{equa:qform}
B_f(\psi_1,\psi_2)=\int_{\mathcal{K}} \frac{dqdp}{2 \pi \hbar} f(q,p) \scalar{\psi_1}{\eta_{q,p}^{[\nu]}}\scalar{\eta_{q,p}^{[\nu]}}{\psi_2}.
\end{equation}
The definition of an operator ${\bf F}$ from this expression is another question and the procedure depends on the bounded or unbounded character of the function $f$.

\subsubsection{$f$ is bounded}

As long as the function $f(q,p)$ is bounded on $\mathcal{K}$, $B_f$ is also bounded as a sesquilinear form. Then the Riesz lemma shows that there exists a unique bounded operator ${\bf F}$ on $\mathcal{H}$ such that
\begin{equation}
B_f(\psi_1,\psi_2)=\scalar{\psi_1}{{\bf F} \psi_2}.
\end{equation}
This gives a precise meaning to the integral notation for ${\bf F}$. Moreover, the mapping $f \mapsto {\bf F}$ is continuous, when both spaces are equipped with ``natural'' norms, because, using the Cauchy-Schwarz inequality, we have $||{\bf F}|| \le ||f||_{\infty}$.

\subsubsection{$f$ is unbounded}

The situation is more complex.
We can at first define the operator ${\bf F}$ on some subspace $\mathcal{D}({\bf F})$ as
\begin{equation}
\label{equa:defOp}
{\bf F} \psi(x)=\int_{\mathcal{K}} \frac{dqdp}{2 \pi \hbar} f(p,q) \scalar{\eta_{q,p}^{[\nu]}}{\psi} \eta_{q,p}^{[\nu]}(x).
\end{equation}
The domain $\mathcal{D}({\bf F})$ is obtained by imposing the existence of the integral (the integrand must be an $L^1(\mathcal{K})$-function) and we add the constraint ${\bf F}\psi \in \mathcal{H}$. The obtained domain $\mathcal{D}({\bf F})$ is a (possibly dense) subspace of $\mathcal{H}$. Moreover,
\begin{equation}
\forall \psi_1 {\rm, } \psi_2 \in \mathcal{D}({\bf F}) {\rm, } \scalar{\psi_1}{ {\bf F} \psi_2} = \int_{\mathcal{K}} \frac{dqdp}{2 \pi \hbar} f(q,p) \scalar{\psi_1}{\eta_{q,p}^{[\nu]}}\scalar{\eta_{q,p}^{[\nu]}}{\psi_2}.
\end{equation}
In the case of real functions $f$, we obtain symmetric operators ${\bf F}$, thus the problem lies in the existence of self-adjoint extensions (and possible uniqueness).

If the function $f$ is positive (or semi-bounded), the Friedrichs extension solves the problem \cite{reedsimon2}: there exists a unique self-adjoint operator associated to the form (in sense of the first representation theorem) such that the domain of the self-adjoint extension is contained in the domain of the quadratic form \cite{reedsimon2}.

If the function $f$ is completely unbounded the problem of self-adjoint extensions is more subtle. In the following we will encounter this situation more than once; in particular we will recover the already mentioned critical value $\nu=1/2$ for ${\bf H}_{\nu,\beta}$.

To summarize the discussion above, the integral expression \eref{equa:qmap} involving unbounded real functions does not automatically provide self-adjoint operators. In general we have only (densely defined) symmetric sesquilinear forms.
Consequently, in the sequel, we study the correspondence $f \mapsto B_f(.,.)$ defined in \eref{equa:qform}.

\subsection{Some operator expressions}

The aim of this section is to show how the definition of the coherent states as eigenstates of ${\bf A}_{\nu,\beta}$ allows to obtain closed formulae for quantized version of a family of classical functions. Furthermore, we want to investigate the self-adjointness of the resulting operators (when possible).

\subsubsection{ ${\bf A}_{\nu,\beta}$, ${\bf A}_{\nu,\beta}^{\dagger}$ and related operators}

First of all let us define the bounded self-adjoint operator ${\bf Q}$ acting on $\mathcal{H}$ as $({\bf Q}\psi)(x)=x \psi(x)$ and three possible candidates ${\bf P}_{0, \pm 1}$ for the ``momentum operator", all acting as $\psi \to -i \hbar \psi'$ on their respective domain  $\mathcal{D}({\bf P}_{-1})=\{ \psi \in AC(0,L) \, | \, \psi(0)=\psi(L)=0 \}$, $\mathcal{D}({\bf P}_{0})=\{ \psi \in AC(0,L) \, | \, \psi(0)=\psi(L) \}$ and $\mathcal{D}({\bf P}_{+1})=\{ \psi \in AC(0,L) \}$. ${\bf P}_{+1}$ is closed (but not symmetric) and  ${\bf P}_{-1}^\dag={\bf P}_{+1}$, ${\bf P}_{-1}$ is closed symmetric (but not self-adjoint), while ${\bf P}_0$ is self-adjoint \cite{reedsimon, faraut2000}. All of them possess a common symmetric restriction ${\bf P}_\#$ on the domain $\mathcal{D}_A$ defined in \eref{equa:domD}. When restricted to $\mathcal{D}_A$, the operators ${\bf A}_{\nu,\beta}$, ${\bf A}_{\nu,\beta}^\dag$,  ${\bf Q}$, and ${\bf P}_\#$ verify ${\bf A}_{\nu,\beta}=W_{\nu,\beta}({\bf Q}) + i {\bf P}_\#$ and ${\bf A}_{\nu,\beta}^\dag=W_{\nu,\beta}({\bf Q}) - i {\bf P}_\#$.

Now let us pick some $\phi, \psi \in \mathcal{D}_A$. Calculating the scalar product $\scalar{\psi}{{\bf A}_{\nu,\beta}^\dag \phi}$ using the resolution of unity \eref{equa:resol} and taking into account the eigen property of our coherent state we obtain
\begin{equation}
\fl 
\scalar{\psi}{{\bf A}_{\nu,\beta}^\dag \phi}=B_{W(q)-ip}(\phi,\psi)=\int_{\mathcal{K}} \frac{dqdp}{2\pi \hbar} (W_{\nu,\beta}(q)-ip) \scalar{\psi}{\eta_{q,p}^{[\nu]}} \scalar{\eta_{q,p}^{[\nu]}}{\phi}
\end{equation}
Since $\overline{\scalar{\psi}{{\bf A}_{\nu,\beta}^\dag \phi}}=\scalar{{\bf A}_{\nu,\beta}^\dag \phi}{\psi}=\scalar{\phi}{{\bf A}_{\nu,\beta} \psi}$, we also deduce
\begin{equation}
\fl 
\scalar{\phi}{{\bf A}_{\nu,\beta} \psi}=B_{W(q)+ip}(\phi,\psi)=\int_{\mathcal{K}} \frac{dqdp}{2\pi \hbar} (W_{\nu,\beta}(q)+ip) \scalar{\phi}{\eta_{q,p}^{[\nu]}} \scalar{\eta_{q,p}^{[\nu]}}{\psi}
\end{equation}
By exchanging the roles of $\phi$ and $\psi$, adding or subtracting the previous equations and taking into account the expression for $W_{\nu,\beta}$, we obtain the following expressions for all $\phi$, $\psi \in \mathcal{D}_A$
\begin{eqnarray}
 \label{equa:cotQ} \scalar{\phi}{\cot ( \pi {\bf Q} L^{-1}) \psi}&=B_{\cot(\pi q L^{-1})}(\phi,\psi) \\
\label{equa:moment} \scalar{\phi}{{\bf P}_\# \psi}&=B_p(\phi,\psi)
\end{eqnarray}
While \eref{equa:cotQ} indicates that there is a ``natural" self-adjoint operator associated to $B_{\cot(\pi q L^{-1})}(.,.)$, it follows from the previous discussion that the equation \eref{equa:moment} yields different closed extension of ${\bf P}_\#$ (the ${\bf P}_{0, \pm1}$, symmetric, but not all self-adjoint) that are compatible with $B_p(.,.)$. Indeed the operator $\cot(\pi {\bf Q} L^{-1})$ is essentially self-adjoint on $\mathcal{D}_A$, \ie~it possesses a unique self-adjoint extension, while ${\bf P}_\#$ is not essentially self-adjoint on $\mathcal{D}_A$ and therefore different closed extensions (self-adjoint or not) exist. This means that $\mathcal{D}_A$ is ``too small" to specify the particular self-adjoint operator. These examples illustrate the difficulties with searching for self-adjoint operator if $f$ is completely unbounded. Recent results \cite{Grubisic-2010} yielding the representation theorem even for indefinite forms may bring possible ways out at least in some cases.

 {\it Conclusion}: The qualitative lesson of these examples is the central role played by the initial (or ``natural") definition domain of the quadratic form $B_f$ in the case of completely unbounded real function $f$. Either the symmetric operator corresponding to $B_f$ is essentially self-adjoint on that domain and then it exists a natural self-adjoint extension and the problem is solved, either the operator is not essentially self-adjoint and we are addressing the problem of selection of the physically relevant self-adjoint extension (if it exists).

 \subsubsection{The Hamiltonians and related operators}

Always using the properties of ${\bf A}_{\nu,\beta}$ and ${\bf A}_{\nu,\beta}^{\dag}$, we obtain for all $\phi$, $\psi \in \mathcal{D}_H$
\begin{equation}
\scalar{\phi}{ {\bf A}_{\nu,\beta} {\bf A}_{\nu,\beta}^{\dag} \psi}=\int_{\mathcal{K}} \frac{dqdp}{2\pi \hbar} (W_{\nu,\beta}^2(q)+p^2) \scalar{\phi}{\eta_{q,p}^{[\nu]}} \scalar{\eta_{q,p}^{[\nu]}}{\psi}.
\end{equation}
This leads to the following expression for ${\bf H}_{\nu,\beta}^{(S)}={\bf H}_{\nu+1,\beta}$
\begin{equation}
\label{equa:hamil1}
\fl 
\scalar{\phi}{{\bf H}_{\nu+1,\beta} \psi}=\int_{\mathcal{K}} \frac{dqdp}{2\pi \hbar} \left\{ \frac{p^2}{2m}+\frac{\E_0 (\nu+1)^2}{\sin^2 \frac{\pi q}{L}} -2 \E_0 \beta \cot \frac{\pi q}{L} \right\} \scalar{\phi}{\eta_{q,p}^{[\nu]}} \scalar{\eta_{q,p}^{[\nu]}}{\psi},
\end{equation}
for $\phi, \psi \in \mathcal{D}_H$.
As the classical function involved in this integral is bounded from below, we know (Friedrichs extension) that the closure of the quadratic form is associated with the self-adjoint operator, namely the P\"oschl-Teller Hamiltonian ${\bf H}_{\nu+1,\beta}$. Nevertheless, we remark that this derivation has been done with the implicit constraint $\nu \ge 0$, therefore the previous formula does not give access to Hamiltonians with $\nu < 1$. Another formula, valid for all positive values of $\nu$, is obtained at the end of this section, see (\ref{equa:p2m}).

Now, using the domain $\mathcal{D}_H$, we apply the procedure of the previous section to ${\bf A}_{\nu,0}^2$ and ${\bf A}_{\nu,0}^{\dag 2}$. It gives for all  $\phi, \psi \in \mathcal{D}_H$
\begin{eqnarray}
\scalar{\phi}{{\bf A}_{\nu,0}^2 \psi}=\int_{\mathcal{K}} \frac{dqdp}{2\pi \hbar}  (W_{\nu,0}(q)+ip)^2 \scalar{\phi}{\eta_{q,p}^{[\nu]}} \scalar{\eta_{q,p}^{[\nu]}}{\psi} \\
\scalar{\phi}{{\bf A}_{\nu,0}^{\dag 2} \psi}=\int_{\mathcal{K}} \frac{dqdp}{2\pi \hbar}  (W_{\nu,0}(q)-ip)^2 \scalar{\phi}{\eta_{q,p}^{[\nu]}} \scalar{\eta_{q,p}^{[\nu]}}{\psi}
\end{eqnarray}
Adding the two previous equations and by subsequent algebraic manipulating we obtain for all  $\phi$, $\psi \in \mathcal{D}_H$
\begin{equation}
\label{equa:hamil2}
\fl 
\scalar{\phi}{ \left(\frac{1}{2m} {\bf P}_\#^2-\frac{\E_0(\nu+1)^2}{\sin^2 \frac{\pi}{L}{\bf Q}} \right) \psi} =\int_{\mathcal{K}} \frac{dqdp}{2\pi \hbar}  \left\{ \frac{p^2}{2m}-\frac{\E_0(\nu+1)^2}{\sin^2 \frac{\pi q}{L}}\right\} \scalar{\phi}{\eta_{q,p}^{[\nu]}} \scalar{\eta_{q,p}^{[\nu]}}{\psi}.
\end{equation}
Once more the classical function involved in this quadratic form is not bounded from below and the associated differential operator is in the limit circle case at both ends ($\nu \ge 0$), thus different closed self-adjoint extensions exist. 
Moreover, the corresponding symmetric operator is not essentially self-adjoint on $\mathcal{D}_H$ and the situation is the same as for ${\bf P}_\#$ (\ie~no ``natural" answer).

Subtracting \eref{equa:hamil1} (with $\beta=0$) and \eref{equa:hamil2} we obtain
\begin{equation}
\label{equa:hamil3}
\fl 
\forall \phi, \psi \in \mathcal{D}_H, \, \scalar{\phi}{ \frac{1}{ \sin^2 \frac{\pi}{L}{\bf Q} } \psi}=\frac{2\nu+2}{2\nu+3} \int_{\mathcal{K}} \frac{dqdp}{2\pi \hbar} \frac{1}{\sin^2 \frac{\pi q}{L}}  \scalar{\phi}{\eta_{q,p}^{[\nu]}} \scalar{\eta_{q,p}^{[\nu]}}{\psi},
\end{equation}
\ie~a positive quadratic form above defining a self-adjoint operator via Friedrichs extension.
From \eref{equa:hamil1} (with $\beta=0$) and \eref{equa:hamil3} we deduce
\begin{equation}
\label{equa:hamil4}
\fl 
\forall \phi, \psi \in \mathcal{D}_H, \, \scalar{\phi}{\frac{1}{2m} {\bf P}_\#^2 \psi}= \int_{\mathcal{K}} \frac{dqdp}{2\pi \hbar} \left\{ \frac{p^2}{2m}-\frac{(\nu+1)^2}{2\nu+3} \frac{\E_0}{\sin^2 \frac{\pi q}{L}}\right\}  \scalar{\phi}{\eta_{q,p}^{[\nu]}} \scalar{\eta_{q,p}^{[\nu]}}{\psi}.
\end{equation}
The classical function involved in the integral is again completely unbounded and the associated differential operator is in the limit circle case at both ends, \ie~not essentially self-adjoint. However, ${\bf P}_\#^2$ defines a positive form on $\mathcal{D}_H$ that provides eventually the Friedrichs extension. 
%
%

Now, from \eref{equa:hamil3} and \eref{equa:hamil4} we obtain the general classical expression associated to a given P\"oschl-Teller Hamiltonian for all (positive) values of $\nu$ and $\beta$ on the domain $\mathcal{D}_H$
\begin{equation}
\fl 
\scalar{\phi}{{\bf H}_{\nu,\beta} \psi}=\int_{\mathcal{K}} \frac{dqdp}{2\pi \hbar}  \left\{ \frac{p^2}{2m}+ \frac{2\nu-1}{2\nu+3} \frac{\E_0(\nu+1)^2}{\sin^2 \frac{\pi q}{L}q} -2\E_0 \beta \cot \frac{\pi q}{L}\right\} \scalar{\phi}{\eta_{q,p}^{[\nu]}} \scalar{\eta_{q,p}^{[\nu]}}{\psi}.
\end{equation}
When $\nu \ge 1/2$ the function in the integral is bounded from below then we know that the associated self-adjoint operator is unique: this means that Dirichlet boundary conditions are automatically imposed and we recover the result of Gesztesy et {\it al.} \cite{Gesztesy}. On the contrary, when $\nu < 1/2$ the function into the integral is completely unbounded and the operator is in the limit circle at both ends (different self-adjoint versions of this differential operator exist). 


We finish this section by this last example that proves that CS quantization can specify boundary conditions in certain circumstances. We deduce from the equations \eref{equa:hamil1} and \eref{equa:hamil3} and the argument of positivity that the following quadratic form specifies the unique self-adjoint operator (for all $\nu \ge 0$) via Friedrichs extension: 
\begin{equation}
\label{equa:p2m}
\fl 
\forall \phi, \psi \in \mathcal{D}_H, \, \int_{\mathcal{K}} \frac{dqdp}{2\pi \hbar} \frac{p^2}{2m} \scalar{\phi}{\eta_{q,p}^{[\nu]}} \scalar{\eta_{q,p}^{[\nu]}}{\psi}=\scalar{\phi}{ \left( \frac{1}{2m} {\bf P}_\#^2+\frac{\nu+1}{2} \frac{\E_0}{\sin^2 \frac{\pi}{L}{\bf Q}} \right) \psi} .
\end{equation}
We know that the corresponding differential operator is in the limit point case at both ends only if $(\nu+1)/2 \ge 3/4$, \ie~$\nu \ge 1/2$. In the range $0 \le \nu < 1/2$ this operator is in fact in the limit circle case and different possible self-adjoint extensions exist, nonetheless, the identity (\ref{equa:p2m}) implies that the CS quantization allows to choose between these possible self-adjoint extensions the ``natural'' one: namely that one corresponding to Dirichlet boundary conditions.

\subsection{Some lowering symbols}

Because the coherent states are the eigenstates of ${\bf A}_{\nu,\beta}$ we deduce
\begin{equation}
\label{equa:low1}
\left\{
\begin{array}{c}
\scalar{\eta_{q,p}^{[\nu]}}{{\bf A}_{\nu,\beta} \eta_{q,p}^{[\nu]}}=W_{\nu,\beta}(q)+ip \\
\scalar{\eta_{q,p}^{[\nu]}}{{\bf H}_{\nu,\beta} \eta_{q,p}^{[\nu]}}=\frac{p^2}{2m}+\frac{\E_0(\nu+1)^2}{\sin^2 \frac{\pi q}{L}}-2\E_0 \beta \cot \frac{\pi q}{L}
\end{array}.
\right.
\end{equation}
Using \eref{equa:intfornorm} we are able to compute another lowering symbol, namely
\begin{equation}
\label{equa:low2}
\scalar{\eta_{q,p}^{[\nu]}} {\frac{1}{\sin^2 \frac{\pi}{L} {\bf Q}} \eta_{q,p}^{[\nu]}}=\frac{2\nu+2}{2\nu+1} \frac{1}{\sin^2 \frac{\pi q}{L}}.
\end{equation}
Finally using \eref{equa:low1} and \eref{equa:low2} we obtain
\begin{equation}
\label{equa:low3}
\scalar{\eta_{q,p}^{[\nu]}} {\frac{1}{2m} {\bf P}_\#^2 \eta_{q,p}^{[\nu]}}=\frac{p^2}{2m}+ \frac{1}{2\nu+1} \frac{\E_0(\nu+1)^2}{\sin^2 \frac{\pi q}{L}}.
\end{equation}

\section{Asymptotic behavior - harmonic oscillator limit}

In this section we assume $\beta=0$. We want to study the limit $L \to \infty$, but in a symmetric way, so we introduce a translated version of our Hilbert space $\mathcal{H}_{T}=L^2([-L/2,L/2],dx)$: the transformation of all previous formulae is straightforward. The Hilbert space limit is that of the particle on the full line. First we notice that
\begin{equation}
\fl 
W_{\nu}(x)=-\frac{\hbar \pi}{L} (\nu+1) \cot \left( \frac{\pi}{L}(x+L/2) \right) = \frac{\hbar \pi}{L} (\nu+1) \tan \left( \frac{\pi x}{L}  \right) \simeq_{L \to \infty} \frac{(\nu+1) \hbar \pi^2 x}{L^2}
\end{equation}
Since the linear behavior of the superpotential corresponds to the case of the harmonic potential, we can guess that there exists an intermediate domain of $L$-values where we can find some features of the harmonic Hamiltonian.

Our coherent states $\ket{\eta_{q,p}^{[\nu]}}$ are defined as
\begin{equation}
\eta_{q,p}^{[\nu]}(x)= N_{\nu}(q) \sin^{\nu+1} \left( \frac{\pi}{L}(x+L/2) \right) \exp \left( \frac{W_{\nu}(q)+ip}{\hbar} x\right)
\end{equation}
with
\begin{equation}
N_{\nu}(q)= \frac{2^{\nu+1} |\Gamma(\nu+2+i (\nu+1) \tan \frac{\pi}{L}q)|}{\sqrt{L} \sqrt{ \Gamma(2\nu+3)}} .
\end{equation}
First we can rewrite $\eta_{q,p}^{[\nu]}(x)$ as
\begin{equation}
\eta_{q,p}^{[\nu]}(x)= N_{\nu}(q) \exp \left( (\nu+1) \ln \sin \left( \frac{\pi}{L}(x+L/2) \right) + \frac{W_{\nu}(q)+ip}{\hbar} x \right)
\end{equation}
and then
\begin{equation}
\eta_{q,p}^{[\nu]}(x)= N_{\nu}(q) \exp \left( (\nu+1) \ln \cos \left( \frac{\pi x}{L} \right) + \frac{W_{\nu}(q)+ip}{\hbar} x\right)
\end{equation}
We deduce the behavior for large values of $L$
\begin{equation}
\eta_{q,p}^{[\nu]}(x) \simeq_{L \to \infty} N_{\nu}(q) \exp \left( - \frac{(\nu+1)\pi^2 x^2}{2 L^2}  + \frac{(\nu+1) \pi^2 q x}{L^2}+\frac{ip x}{\hbar} \right)
\end{equation}
with
\begin{equation}
N_{\nu}(q) \simeq_{L \to \infty} \frac{2^{\nu+1} \Gamma(\nu+2)}{\sqrt{L} \sqrt{\Gamma(2\nu+3)}}
\end{equation}
Then our coherent states degenerate into harmonic coherent states, while the complete asymptotic behavior corresponds to a plane wave
\begin{equation}
\eta_{q,p}(x) \simeq_{L \to \infty} \frac{2^{\nu+1} \Gamma(\nu+2)}{\sqrt{L} \sqrt{\Gamma(2\nu+3)}} e^{i p x/\hbar}
\end{equation}

\section{Conclusion}

We analyzed the mathematical features of our coherent states, in particular the CS quantization of some unbounded real functions, studying the existence and uniqueness of (possible) self-adjoint operators associated to these functions. We exhibited some interesting, generally expected, qualitative properties (not restricted to these specific examples):
\begin{itemize}
\item When the classical real function involved in the quadratic form is semi-bounded, a unique self-adjoint operator is associated to the form via Friedrichs extension. This means that CS quantization is  (sometimes) able to select a unique self-adjoint operator in situations where many possible self-adjoint extensions exist (in this sense, the CS quantization includes implicitly the boundary conditions).
\item When the classical function involved in the quadratic form is completely unbounded, the situation is more difficult due to the lack of representation theorems. We can consider corresponding symmetric operator, but
\begin{itemize}
\item Either that operator is essentially self-adjoint on the domain and the unique self-adjoint extension is available,
\item Either that operator is not essentially self-adjoint and finding a ``natural'' self-adjoint operator corresponding to that form is generally a well known problem where additional physical information is needed for selecting particular self-adjoint extension. Nonetheless, the CS quantization allows in some cases select the ``natural'' extension, see comments at (\ref{equa:p2m} ).
\end{itemize}
\end{itemize}
These results also illustrate the strong limitations of formal manipulations only based on Dirac formalism (when unbounded functions are involved). 

Finally, we showed that our CS degenerate into usual harmonic CS in the limit $L \to \infty$, constituting a continuous transition between the framework of a particle trapped on an interval and that of a free particle on the full real line.

\section*{Acknowledgments}
We want to acknowledge Prof. J.P. Gazeau for fruitful and numerous discussions. P.S. appreciates the support of the Czech Ministry of Education, Youth, and Sports within the project LC06002, the GACR grant No. P203/11/0701 and the Grant Agency of the Czech Technical University in Prague, grant No. SGS OHK4-010/10.

\appendix

\section{Norm Formula}
Taking $p=\nu+2+i \frac{a}{2 \pi}$, $q=\nu+2-i \frac{a}{2 \pi}$ in the integral relation \cite{gradshtein1}
\begin{equation}
\fl 
\int_0^{\pi/2} dx \cos^{p+q-2} x \cos (p-q)x = \frac{\pi}{2^{p+q-1} (p+q-1) {\rm B}(p,q)}, \quad p+q > 1, 
\end{equation}
and changing the variable we obtain
\begin{equation}
\fl 
\forall \nu > -3/2, \ \int_0^{1/2} dx \cos^{2\nu+2} \pi x \cosh a x = \frac{2^{-(2\nu+3)}} {(2\nu+3) {\rm B}(\nu+2+i \frac{a}{2 \pi},\nu+2-i \frac{a}{2 \pi})}
\end{equation}
Then by parity
\begin{equation}
\int_{-1/2}^{1/2} dx \cos^{2\nu+2} (\pi x) e^{a x} = \frac{2^{-(2\nu+2)}} {(2\nu+3) {\rm B}(\nu+2+i \frac{a}{2 \pi},\nu+2-i \frac{a}{2 \pi})}
\end{equation}
and
\begin{equation}
\int_0^1 dx \cos^{2\nu+2} (\pi x -\pi/2) e^{a x} = \frac{2^{-(2\nu+2)} e^{a/2}} {(2\nu+3) {\rm B}(\nu+2+i \frac{a}{2 \pi},\nu+2-i \frac{a}{2 \pi})}.
\end{equation}
Developing the ${\rm B}(p,q)$ function in terms of $\Gamma$ functions, we obtain finally
\begin{equation}
\fl 
\forall a \in \mathbb{C}, \ \forall \nu>-3/2 {\rm, } \int_0^1 \sin^{2\nu+2}( \pi x ) e^{a x} dx= \frac{\Gamma(2\nu+3) e^{a/2}}{4^{\nu+1} \Gamma(\nu+2+i \frac{a}{2\pi}) \Gamma(\nu+2-i \frac{a}{2\pi})}.
\end{equation}

\section{Integral for the resolution of unity}
We start from a well-kown Fourier transform \cite{gradshtein2}
\begin{equation}
\fl 
\forall k \in \mathbb{R}, \ \forall \nu >-1, \ \int_{\mathbb{R}} \frac{e^{-i k x}}{\cosh^{2\nu+2}  (x)} \frac{dx}{2 \pi}=\frac{4^{\nu} \Gamma(\nu+1-i \frac{k}{2}) \Gamma(\nu+1+i \frac{k}{2})}{\pi \Gamma(2\nu+2)}.
\end{equation}
Using the inverse Fourier transform we find
\begin{equation}
\fl 
\forall x \in \mathbb{R} {\rm, } \, \forall \nu >-1 {\rm, } \int_{\mathbb{R}} \frac{4^{\nu} \Gamma(\nu+1-i \frac{k}{2}) \Gamma(\nu+1+i \frac{k}{2})}{\pi \Gamma(2\nu+2)} e^{i k x} dk= \frac{1}{\cosh^{2\nu+2} x}
\end{equation}
Due to the uniqueness of analytical extension, we can extend the previous equality for $x \in \mathbb{C}$ with the constraint $-\pi/2 < \Im (x) <\pi/2$. Taking $u=i x$ as a new variable, we obtain
\begin{equation}
\fl 
\forall u \in \mathbb{C} {\rm, }\,  |\Re(u)|< \pi/2 {\rm, }\,  \forall \nu >-1 {\rm, } \int_{\mathbb{R}} \frac{4^{\nu} \Gamma(\nu+1-i \frac{k}{2}) \Gamma(\nu+1+i \frac{k}{2})}{\pi \Gamma(2\nu+2)} e^{k u} dk= \frac{1}{\cos^{2\nu+2} u}.
\end{equation}
By a change of variable, we have finally
\begin{equation}
\fl 
\forall x {\rm, } \,  0<x<1 {\rm, }\,  \forall \nu > -1 {\rm, }  \int_{\mathbb{R}} \frac{4^{\nu} |\Gamma(\nu+1-i \frac{k}{2 \pi})|^2 e^{-k/2}}{\pi^2 \Gamma(2\nu+2)} e^{k x} dk= \frac{1}{\sin^{2\nu+2} \pi x}.
\end{equation}

\section*{References}


\begin{thebibliography}{99}

\bibitem{PT} P\"oschl G  and Teller E 1933 {\it Z. Physik} {\bf 83} 143.
\bibitem{RM} Rosen N and Morse P M  1932 {\it Phys. Rev.} {\bf 42} 210.
\bibitem{Tomak}  Yildirim H and Tomak M  2005 {\it Phys. Rev. B} {\bf 72} 115340.
\bibitem{Wang} Wang G, Guo Q, Wu L and Yang X  2007 {\it Phys. Rev. B} {\bf 75}205337.
\bibitem{Gend} Gendenshtein L  1983 {\it JETP Lett.} {\bf 38} 356.
\bibitem{Book} Cooper F, Khare A and Sukhatme U P 2002 {\it Supersymmetry in Quantum Mechanics} (World Scientific Publishing Company, Singapore).
\bibitem{Bala1} Balantekin A B 1998 {\it Phys. Rev. A} {\bf57} 4188
\bibitem{Fukui} Fukui T and Aizawa N 1993 {\it Phys. Lett. A} {\bf180} 308
\bibitem{Fatyga} Fatyga B W, Kosteleck\'y V A, Nieto M M and Truax D R  1991 {\it Phys. Rev. D}  {\bf 43} 1403
\bibitem{Shreecharan} Shreecharan T, Panigrahi P K and Banerji J 2004 {\it Phys. Rev. A} {\bf 69} 012102
\bibitem{Crawford} Crawford M G A and Vrscay E R 1998 {\it Phys. Rev. A} {\bf 57} 106
\bibitem{Alex} Alexio A and Balantekin A B 2007 {\it J. Phys. A} {\bf 40} 3463
\bibitem{Gaz} Antoine J P, Gazeau J P, Monceau P, Klauder J R and Penson K A  2001 {\it J. Math. Phys.} {\bf 42} 2349
\bibitem{Kinani} El Kinani A H and Daoud M  2001 {\it Phys. Lett. A} {\bf 283} 291
\bibitem{Fernandez} Fernandez D J, Hussin V and Rosas-Ortiz O 2007 {\it J. Phys. A: Math. Theor.}  {\bf 40} 6491
\bibitem{Cruz} Cruz y Cruz S, Kuru S, Negro J 2008  {\it Phys. Lett. A} {\bf 372} 1391
\bibitem{reedsimon} Reed M and Simon B 1972 {\it Methods of Modern Mathematical Physics, I. Functional Analysis} (Academic Press, New York)
\bibitem{HJP2010} Bergeron H, Gazeau J P, Siegl P and Youssef A 2010, {\it EPL} {\bf 92} 60003
\bibitem{Herve} Bergeron H and Valance A 1995 {\it J. Math. Phys.} {\bf 36} 1572
\bibitem{Compean2008} Compean C B and Kirchbach M, 2008 {\it arxiv:quant-ph/0509055v2}
\bibitem{Gesztesy} Gesztesy F and Kirsch W  1985 {\it J. Rein. Ang. Math.} {\bf 362} 28
\bibitem{Mielnik} Mielnik B and Rosas-Ortiz O 2004 {\it J. Phys. A: Math. Gen.} {\bf 37} 10007
\bibitem{Andrianov} Andrianov A A and Cannata F 2004 {\it J. Phys. A: Math. Gen.} {\bf 37}  10297
\bibitem{csbooks} Klauder J R and Skagerstam B S 1985 {\it Coherent states, applications in physics and mathematical physics} (World scientific, Singapore)
\bibitem{alienglis2005} Ali S T and Engli\v{s} M  2005  {\it Rev. Math. Phys.} {\bf 17} 391-490
\bibitem{gazeau2009} Gazeau J P 2009 \textit{Coherent States in Quantum Physics} (Wiley-VCH)
\bibitem{faraut2000} Bonneau G, Faraut J, Valent G 2000 {\it Am. J. Phys.} {\bf 69} (3) 322
\bibitem{reedsimon2} Reed M and Simon B  1975 {\it Methods of Modern Mathematical Physics, II. Fourier Analysis, Self-Adjointness}  (Academic Press, New York)
\bibitem{Grubisic-2010} Grubisic L., Kostrykin V., Makarov K. A. and  Veselic, K., 2010 {\it arxiv:1003.1908}
\bibitem{gradshtein1} Gradshteyn-Ryznik p. 375
\bibitem{gradshtein2} Gradshteyn-Ryznik p. 520

\end{thebibliography}
\end{document}